\documentclass[     ,final              ]{aipproc}%
\usepackage{amsmath}%
\usepackage{amsfonts}%
\usepackage{amssymb}%
\usepackage{graphicx}
\input{aipcheck}
\layoutstyle{6x9}
\begin{document}
\title{Charge Fluctuation of Dust Grain and Its Impact on Dusty-Acoustic
Wave Damping}

\author{B. Atamaniuk}{
  address={Institute of Fundamental Technological Research, Polish Academy of
  Sciences\\
00-049 Warsaw \'{S}wietokrzyska 21 POLAND} }
\author{K. \.{Z}uchowski}{
  address={Institute of Fundamental Technological Research, Polish Academy of
  Sciences\\
00-049 Warsaw \'{S}wietokrzyska 21 POLAND}}

\begin{abstract}
We consider the influence of dust charge fluctuations on damping of the
dust-ion-acoustic waves. It is assumed that all grains have equal masses but
charges are not constant in time - they may fluctuate in time. The dust
charges are not really independent of the variations in the plasma potentials.
All modes will influence the charging mechanism, and feedback will lead to
several new interesting and unexpected phenomena. The charging of the grains
depends on local plasma characteristics. If the waves disturb these
characteristic, then charging of the grains is affected and the grain charge
is modified, with a resulting feedback on the wave mode. In the case
considered here, when the temperature of electrons is much greater than the
temperature of the ions and the temperature of electrons is not great enough
for further ionization of the ions, we show that attenuation of the acoustic
wave depends only on one phenomenological coefficient

\end{abstract}
\maketitle


\section{Introduction}

Dusty plasma represent the most general form of space, laboratory and
industrial plasmas. Dusty plasma are conglomerations of the ions, electrons
and neutral particles. These large particles, to be called grains, have atomic
numbers $Z_{d}$ in the range of $10^{4}-10^{6}$ and their mass $m_{d}$ can be
equal to $10^{6}$ of the \ proton mass or even much more. In the considered
dusty plasmas, the size of grains is small compared with average distance
between the grains. The ratio of charge to mass for a given component of
plasma determines its dynamics. We note that for dusty plasmas, ratio of
electrical charges of grains to their masses is usually much smaller than in
the case of multispecies plasmas with negative ions and hence, here comes the
first of the crucial differences between multispecies plasmas with negative
ions and dusty plasmas. If it is assumed that all grains have equal masses and
charges steady in time, therefore the dust-ion-acoustic and dust-acoustic
dispersion relations are obtained on the basis of fluid \cite{Shukla92},
\cite{Shukla92a} or kinetic \cite{Turski99} models. For simplicity we have
assumed, that all grains have equal masses and charges, but charges are not
constant in time - they may fluctuate in time. The dust charges are not really
independent of the variations in the plasma potentials. Here, even in the
fluid theory, appear\ the crucial differences between the ordinary
multispecies plasmas and the dusty plasmas. All modes will influence the
charging mechanism, and feedback will lead to several new interesting and
unexpected phenomena. The charging of the grains depends on local plasma
characteristics. If the waves disturb these characteristic, then charging of
the grains is affected and the grain charge is modified, with a resulting
feedback on the wave mode.

\section{Fluctuation of dust grains in dusty plasmas}

We consider the parallel electrostatic modes in an unmagnetized plasma when
the temperature of electrons $T_{e}$ is much greater than the temperature of
ions $T_{i}$: $T_{e}\gg T_{i}$. In such simplified situations, fluctuations in
time of the number density of electrons $\delta n_{e}$ can occur due to the
grains of the dust loosing or picking up some electrons. As a result of
fluctuating dust charges in dusty plasmas, many new problems\ can appear which
are in partly treatment by Verheest \cite{Verheest00}. We also assume that the
mass of grains with fluctuating charges may be approximated by constant
values. In this case the continuity equations for specimens of dusty plasmas
can be written in the form:%
\[
\partial n_{d}/\partial t+\partial(n_{d}u_{d})/\partial x=0,
\]%
\begin{equation}
\partial n_{i}/\partial t+\partial(n_{i}u_{i})/\partial x=0, \tag{2.1}%
\label{2.1}%
\end{equation}%
\[
\partial n_{e/}\partial t+\partial(n_{e}u_{e})/\partial x=S_{e}.
\]

\bigskip Due to the possible fluctuations of the dust charges we can express
the conservation of charge in the dusty plasma by:%

\begin{equation}
\frac{\partial}{\partial t}\left(  -n_{e}e+n_{d}q_{d}+n_{i}e\right)
+\frac{\partial}{\partial x}\left(  -n_{e}eu_{e}+n_{d}q_{d}u_{d}+n_{i}%
eu_{i}\right)  =0, \tag{2.2}\label{2.2}%
\end{equation}
where $q_{d}$ is the charge of grain of dust. This can be rewritten with the
help of the continuity equation (2.1 - 2.3) as:%

\begin{equation}
n_{d}\left(  \frac{\partial}{\partial t}+u_{d}\frac{\partial}{\partial
x}\right)  q_{d}=e\,S_{e}. \tag{2.3}\label{2.3}%
\end{equation}
On the other hand, the charge of grain of dust fluctuation is given by:
\begin{equation}
\frac{dq_{d}}{dt}=\left(  \frac{\partial}{\partial t}+u_{d}\frac{\partial
}{\partial x}\right)  q_{d}=I_{i}\left(  n_{i,}q_{d}\right)  +I_{e}\left(
n_{e},q_{d}\right)  , \tag{2.4}\label{2.4}%
\end{equation}
where $I_{i}\left(  n_{i},q_{d}\right)  $ and $I_{e}\left(  n_{e}%
,q_{d}\right)  $ are the ionic and electronic charging current, respectively.
When we combine (2. 3) and (2.4), we get
\begin{equation}
eS_{e}=n_{d}I_{e}\left(  n_{e},q_{d}\right)  +n_{d}I_{i}\left(  n_{i,}%
q_{d}\right)  . \tag{2.5}\label{2.5}%
\end{equation}
In equilibrium dusty plasma, the total charging current vanishes:
\begin{equation}
I_{i0}+I_{e0}=0, \tag{2.6}\label{2.6}%
\end{equation}
where $I_{i0}$ and $I_{e0}$ denotes the equilibrium charging current for ions
and electrons, respectively. Therefore we can expand (2.5) as a function of
$n_{e}$, $q_{d}$ and $n_{d}$ using (2.6) and \ hence in linear approximation
for $S_{e}$ vanishing at equilibrium, it is given by:
\begin{equation}
S_{e}=-\nu_{e}\delta n_{e}-\mu_{e}\delta q_{d}, \tag{2.7}\label{2.7}%
\end{equation}
where $\nu_{e}$, $\mu_{e}$ denotes charging fluctuation coefficients while
$\delta n_{e}$ \ and $\delta q_{d}$ denotes fluctuation electron number
density and \ fluctuation charges of grains from their equilibrium values respectively.

\section{Dumping of Dust-Ion-Acustic Wave}

Now we add to the continuity equation \ref{2.1} some dispersion relations for
ideal dusty plasma, when the fluctuation of the charge of the dust grain is
absent. To determine dispersion relation we used the linear response theory
\cite{Schram91}, \cite{Nishikawa}.

In Fourier representation we have
\begin{equation}
q_{\alpha}(k,\omega)\delta n_{\alpha}=k^{2}\chi_{\alpha}(k,\omega
)\phi(k,\omega) \tag{3.1}\label{3.1}%
\end{equation}

where $\delta n_{\alpha}$- number density fluctuation of the $\alpha$
components of dusty plasma, $\chi_{\alpha}$- susceptibility. Dispersion
relation for the ideal dusty plasma is given by%
\begin{equation}
\varepsilon(k,\omega)=\varepsilon_{0}\left(  1+\sum\limits_{_{\alpha}}%
\chi_{\alpha}(k,\omega)\right)  . \tag{3.2}\label{3.2}%
\end{equation}
Then, using Poisson equation with global charge neutrality after linearization
and Fourier transform we received:%
\begin{equation}
\delta q_{d}(k,\omega)=\frac{-\nu_{e}\frac{e}{n_{d0}}\left(  i\omega+\frac
{e}{n_{d0}}\mu_{e}\right)  \delta n_{e}(k,\omega)}{\omega^{2}+\left(  \frac
{e}{n_{d0}}\mu_{e}\right)  ^{2}} \tag{3.3}\label{3.3}%
\end{equation}
$n_{d0}$ denotes the equilibrium number density of the dust. Next \ for
$\alpha=e$ we received dispersion relation for acoustic wave, which take into
account fluctuation of the grain charge:%
\begin{equation}
1+\sum\limits_{_{\alpha}}\chi_{\alpha}(k,\omega)=\frac{-\nu_{e}\frac{e}%
{n_{d0}}\left(  i\omega+\frac{e}{n_{d0}}\mu_{e}\right)  \chi_{e}%
(k,\omega)n_{d0}}{e\left(  \omega^{2}+\left(  \frac{e}{n_{d0}}\mu_{e}\right)
^{2}\right)  } \tag{3.4}\label{3.4}%
\end{equation}

If $\chi_{e}\approx\frac{1}{k^{2}\lambda_{De}^{2}};$ $\chi_{i}\approx
\frac{\omega_{pi}^{2}}{\omega^{2}}$ and $\chi_{d}\approx\frac{\omega_{pd}^{2}%
}{\omega^{2}}$ where $\lambda_{D\alpha}$, $\omega_{pd}^{2}$ Debye length and
plasma frequency for $\alpha$component respectively. In our case the $\nu_{e}$
and $\frac{e}{n_{d0}}\mu_{e}$ are smaller than%

\begin{equation}
\omega_{0}=\sqrt{\frac{k^{2}\lambda_{De}^{2}\omega_{pi}^{2}}{1+k^{2}%
\lambda_{De}^{2}}+k^{2}\frac{k_{B}T_{i}}{m_{i}}} \tag{3.5}\label{3.5}%
\end{equation}
and the dispersion relation for DIAW waves is given by%
\begin{equation}
\omega=\omega_{0}+i\frac{k^{2}\lambda_{De}^{2}\omega_{pi}^{2}}{2\omega_{0}%
^{2}\left(  1+k^{2}\lambda_{De}^{2}\right)  }\nu_{e}. \tag{3.6}\label{3.6}%
\end{equation}
For $k\longrightarrow0$, we have%

\begin{equation}
\omega=\omega_{0}+i\nu_{e}\frac{\lambda_{De}^{2}}{2\left(  \lambda_{Di}%
^{2}+\lambda_{De}^{2}\right)  }\approx k\lambda_{De}\omega_{pi}-\frac{i\nu
_{e}}{2} \tag{3.7}\label{3.7}%
\end{equation}

This equation describes the dumped dust-ion-acoustic waves \ including the
charge fluctuation. In our approximation: \ $T_{e}>>T_{i}$, $\omega_{0}%
>>\nu_{e}$and $\omega_{0}>>\frac{e}{n_{d0}}\mu_{e}$ then the dumping of
dust-ion-acoustic waves is dependent on the one phenomenological parameter
$\nu_{e}$.

\section{Conclusions}

The paper deals with a small dust charge fluctuations. In the case considered
here , when the temperature of electrons is much greater than the temperature
of the ions: $T_{e}>>T_{i}$ and $T_{e}$ is not great enough for further
ionization of the ions, we show that attenuation of the acoustic wave depends
only on one phenomenological coefficient $\nu_{e}$. The value of this
coefficient depends mainly on the temperature of electrons.


\begin{theacknowledgments}
This research is supported by KBN grant 2PO3B-126- 24
\end{theacknowledgments}



\bibliographystyle{aipproc}
\bibliography{sample}

\IfFileExists{\jobname.bbl}{}  {\typeout{}  \typeout{******************************************}  \typeout{** Please run
"bibtex \jobname" to optain}  \typeout{** the bibliography and then re-run
LaTeX}  \typeout{** twice to fix the references!}  \typeout{******************************************}  \typeout{}  }

\end{document}